# Reversible structural transformations in supercooled water from 135 to 245 K

Loni Kringle†, Wyatt A. Thornley†, Bruce D. Kay*, and Greg A. Kimmel*

Physical Sciences Division, Pacific Northwest National Laboratory, P.O. Box 999, Richland, WA 99352

†L. Kringle and W. Thornley contributed equally to this work
*Authors to whom correspondence should be addressed. Electronic addresses: gregory.kimmel@pnnl.gov and bruce.kay@pnnl.gov

Water has many anomalous properties compared to "simple" liquids, and these anomalies are typically enhanced in supercooled water.[1-3] While numerous models have been proposed, including the liquid-liquid critical point,[4,5] the singularity-free scenario,[6] and the stability limit conjecture,[1] a molecular-level understanding remains elusive. The main difficulty in determining which, if any, of these models is correct is the limited amount of data in the relevant temperature and pressure ranges. For water at ambient pressures, which is the focus of this work, data is largely missing from 160 – 232 K ("No Man's Land") due to rapid crystallization.[2,3] Whether rapid crystallization is just an experimental obstacle, or a fundamental problem signaling the inability of water to thermally equilibrate prior to crystallization is also a major unanswered question.[5,7] Here, we investigate the structural transformations of transiently-heated, supercooled water with nanosecond time resolution using infrared vibrational spectroscopy. The experiments demonstrate three key results. First, water's structure relaxes from its initial configuration to a "steady-state" configuration prior to the onset of crystallization over a wide temperature range. Second, water's steady-state structure can be reproduced by a linear combination of two, temperature-independent structures that correspond to a "high-temperature liquid" and a "low-temperature liquid." Third, the observed structural changes are reversible over the full temperature range. Taken together, these results show that supercooled water can equilibrate prior to crystallization for temperatures from the homogeneous nucleation temperature, $T_H$ ~232 K,[3] down to the glass transition temperature ($T_g$ ~ 136 K). Second, the results provide support for the hypothesis that supercooled water can be described as a mixture of two, structurally-distinct, interconvertible liquids from 135 K to 245 K.[5,8-18]

Figure 1a shows a schematic of the experimental approach.[19,20] Water films ($H_2O$ with 10% HDO) with typical thicknesses of ~15 nm (50 molecular layers) were adsorbed on a Pt(111) crystal in ultrahigh vacuum at 70 K. The sample was heated with nanosecond laser pulses that produced heating rates of ~$10^{10}$ K/s. For each heat pulse, the water layer spent ~3 ns near the maximum temperature, $T_{max}$, before rapid cooling (~$5 \times 10^9$ K/s) returned the sample to 70 K where water was kinetically frozen (Fig. 1b). Changes in the water induced by the heat pulse were measured with infrared (IR) spectroscopy at 70 K. A "stop-action-movie" of the structural evolution was generated by repeated application of the heat pulses. Details are presented in the Methods section A.

Figure 1c shows the evolution of the IR spectra for two water films transiently-heated to $T_{max}$ = 215 K. The films were prepared with two different initial structures. A "low-temperature liquid" (LTL) was produced by annealing the water film at or near the glass transition temperature, $T_g \approx 136$ K, for ~100 s (Fig 1c(i), blue line). A quenched "high-temperature liquid" (HTL) was produced by transiently-heating to 297 K for one or more pulses (Fig. 1c(i), red line). The IR spectrum of crystalline ice (CI) is also shown for comparison (Fig. 1c(i), green line). Because the IR spectra are sensitive to the hydrogen-bonding configuration of the water molecules within the films,[21,22] the spectra indicate that there are significant structural differences between HTL, LTL and CI.

Upon transient-heating to 215 K, the water IR spectra smoothly evolve as the number of heat pulses, $N_p$, increases (Figs. 1c(ii), 1c(iii), and Extended Data Fig. 1). The spectra also display an isosbestic point at 3,418 cm$^{-1}$. Independent of the starting configuration (HTL or LTL), the spectra observed after transiently-heating to 215 K and just prior to the onset of measurable crystallization (e.g. $N_p = 670$) are the same (Fig. 1c(i), gray and black lines). Further pulsed heating leads to crystallization of the water films (Extended Data Fig. S2). Evolution of the IR spectra from HTL or LTL to a common intermediate state was also observed for other temperatures (Extended Data Fig. 3). All the intermediate spectra can be fit to a linear combination of the LTL, HTL and CI spectra (Extended Data Fig. 4). Because each heat pulse lasted ~3 ns near the temperature maximum, the observed spectral evolution was occurring in the nano- and micro-second regime.

For $T_{max} = 215$ K, the fraction of the water that can be assigned to LTL, HTL and CI ($f_{LTL}$, $f_{HTL}$, and $f_{CI}$, respectively) versus $N_p$ are shown in Figure 2. The procedure for determining these fractions from the IR spectra is discussed in the Methods section B. For both initial configurations, $f_{HTL}$ asymptotically approaches ~0.62 prior to crystallization (red triangles and red open triangles). Similarly, $f_{LTL}$ approaches ~0.38 (blue squares and blue open squares). We note that there are systematic differences in the relaxation kinetics when starting from the LTL and HTL states, with the conversion from the HTL being notably faster (see Fig. 2 and Extended Data Fig. 3). These differences in the kinetics will be the focus of future investigations.

Previous approaches for investigating water below $T_H$ typically involve rapidly cooling small water droplets once[12] or using model aqueous systems that do not crystallize.[9,23] One strength of the pulsed-heating experiments is the ability to bring a sample of water into the supercooled region multiple times. In this case, changing the temperature during an experiment allows one to explore the reversibility of the structural changes such as those shown in Figures 1 and 2. Figure 3a shows $f_{HTL}$ and $f_{LTL}$ versus $N_p$ for water where the temperature was repeatedly cycled between 215 K for 100 pulses and then 252 K for 10 pulses. (A subset of the corresponding IR spectra is shown in Extended Data Fig. 5.) The key observation is that each time the pulsed-heating temperature is changed, the water structure quickly evolved to a new "steady state" characteristic of that temperature. Note that for this experiment the temperatures were chosen to be above and below $T_H$.

Reversible structural changes were also observed for water when the temperature was cycled between 135 K (isothermal heating for 130 s) and 215 K (pulsed heating for $1 \leq N_p \leq 400$) (Fig. 3b and Extended Data Fig. 5b). Because the relaxation kinetics were slower when starting with LTL, more heat pulses were required for the water film to reach its steady-state configuration at 215 K in Fig. 3b than when the starting configuration was HTL (Fig. 3a). As a result, the onset for measurable crystallization occurs after only two temperature cycles for the experiment shown in (Fig 3b). However, the key observation is that for both experiments, the structural changes seen at 215 K can be reversed by simply changing the temperature. More examples of reversible structural transformations for transiently-heated, supercooled water are shown in Extended Data Fig. 6. The results in Figures 1 – 3 show that the observed changes in the water are not irreversible structural transformations associated with crystallization or crossing a spinodal.[1,7,24] Instead, the data indicate that deeply supercooled water thermally equilibrates prior to crystallization.

The results suggest that the IR spectra observed just prior to crystallization reflect the steady-state structure of supercooled water at that temperature. We find that at all temperatures, these steady-state spectra, $S(T, \omega)$, can be fit as a linear combination of the HTL and LTL spectra ($S_{HTL}(\omega)$ and $S_{LTL}(\omega)$, respectively):

$$S(T, \omega) = f_{HTL}(T) \cdot S_{HTL}(\omega) + f_{LTL}(T) \cdot S_{LTL}(\omega), \qquad (1)$$

where $\omega$ is the frequency. Figure 4 shows $f_{HTL}(T)$ determined from the average of measurements of individual water films on both Pt(111) and graphene with initial configurations of LTL and HTL (see also Extended Data Fig. 7). The solid line in Figure 4 shows a fit to the data using the logistic function,[18] $g_l(T) = (1 + exp(-(T - T_0)/\Delta T))^{-1}$, with $T_0 = 210 \pm 2$ K and $\Delta T = 8.5$ K. For $T <$ ~170 K, the water is essentially 100% LTL, but at higher temperatures $f_{HTL}$ increases until it approaches 1 for $T >$ 245 K.

We believe the results described above, which are for water films with coverages of 50 monolayers (i.e. ~15 nm thick), are representative of the behavior of bulk water and are not strongly influenced by the nanoscale thickness of the films. Because the structure of water at interfaces converges to the bulk structure within ~3 – 4 monolayers,[25,26] we expect that most of the water experiences a bulk environment. For water adsorbed on Pt(111) or graphene/Pt(111), $f_{HTL}(T)$ is the same within experimental error (Extended Data Fig. 7) suggesting that the substrate did not play a significant role. Capping the water films with a decane layer also did not appreciably change the results (Extended Data Fig. 8). We also investigated the structural evolution versus $N_p$ in water films with coverages from 25 to 100 ML and the structural changes versus temperature were similar in all cases. Finally, the water films are flat, so the internal pressure is not raised due to curvature effects. Collectively, these results indicate that the experiments are germane for understanding the properties of bulk water.

The ability to decompose the "steady-state" IR spectra of supercooled water into a linear combination of two components is consistent with two-component models such as the singularity-free scenario and the liquid-liquid critical point model.[2-4,6] The characteristics of the "HTL" and "LTL" IR spectra in the OH-stretching region (e.g. Fig. 1) are also consistent with the expectations from two-component models. Specifically, the "LTL" spectrum is red-shifted relative to the "HTL" spectrum indicating stronger hydrogen-bonding, i.e. bonding associated with more "ice-like" configurations.[21,22] IR spectra calculated from classical molecular dynamics simulations of supercooled water and a procedure for relating the local electric field to the IR spectra agree qualitatively with the IR spectra observed here.[22]

The observation that $f_{HTL}(T)$ is ~1 for T > 245 K is related to limitations of the experimental technique. Due to the finite cooling rate, the experiments are not sensitive to changes in the structure at high temperatures where the relaxation rates are fast. At temperatures above ~280 K, water equilibrates during a single pulse near $T_{max}$ and can maintain thermal equilibrium as the water cools until the relaxation rate slows to the point where it cannot keep up with the rapid cooling. This "freeze-in" temperature upon cooling occurs at ~245 K. This explanation is consistent with the observation that the number of pulses required for the water structure to equilibrate decreases as the temperature increases and is ~1 for $T_{max}$ > 245 K. An example of this fast relaxation is shown in Figure 3a, where $f_{HTL}$ increases from ~0.6 to 1 in two pulses for $T_{max}$ = 252 K. To explore the (hyper-quenched) structure of water at higher temperatures would require significantly higher cooling rates than can be achieved using the current approach. Therefore, our results are not sensitive to the structure of water at higher temperatures and any possible inhomogeneities at ambient temperatures.[3,14,17,27,28]

Two-component models propose that water is an inhomogeneous liquid with regions that have higher density and entropy (a high-density liquid, or "HDL") and regions with lower density and entropy (a low-density liquid, or "LDL").[10,13,14,17] Therefore, higher or lower temperatures favor HDL or LDL, respectively. The range of temperatures, $\Delta T$, over which water switches from being predominantly HDL to predominantly LDL is an important prediction of the various two-component models. In the critical-point-free scenario, the coexistence line between HDL and LDL extends to negative pressures.[3] Therefore, water would have a first-order phase transition (i.e. $\Delta T = 0$) at ~zero pressure, which is inconsistent with the current results. In the LLCP scenario, the first-order phase transition occurs at higher pressures, and $\Delta T$ observed at lower pressures provides information on the proximity to the LLCP.[3,10,14,29,30] In the singularity-free scenario,[6] there is no phase transition, but the amount of "non-ideality" governs the width of the transition between HDL and LDL.

While the results shown in Figure 4 support the basic premise of two-component models, to make quantitative comparisons one needs to assess the amount of HDL and LDL in our two starting configurations, HTL and LTL. Because LTL is prepared by annealing at $T_g$, we assume it is essentially 100% LDL. The temperature-independence of the IR spectra below ~170 K also support this assumption. As noted above, HTL reflects the structure of water at ~245 K. Because this is well within the temperature range where water exhibits anomalous properties, it is likely that our HTL is a mixture of HDL and LDL. Our experiments do not allow us to unambiguously determine the composition of HTL and therefore directly comparing our measured $f_{HTL}(T)$ to various two-component models is a challenge. Furthermore, the wide variations in literature suggested for $f_{HDL}(T)$ make it difficult to decide which, if any, of these could be used to assign the composition of HTL (see Extended Data Fig. 9). However, the observation that the steady-state structure changes rapidly above ~195 K and much more slowly at lower temperatures is independent of this scaling issue and should thus be a useful point for comparison to various water models.

Previous experiments using X-ray, Raman, and IR spectroscopy, and the optical Kerr effect have shown bimodal characteristics for normal and supercooled water above $T_H$.[11,12,31,32] Our results are qualitatively consistent with these experiments and extend them to much lower temperatures. An important distinction is that the earlier measurements were all made at the temperature of interest, while the measurements reported here were all made at 70 K after transient heating to $T_{max}$. Due to the high cooling rates, the quenched state reflects the structure of the water at $T_{max}$ (for $T_{max} < 245$ K) but with reduced thermal broadening due to the lower temperature (70 K) where the IR measurements were performed. Therefore, our experimental results more closely match the "inherent structures" obtained from molecular dynamics simulations where the system is instantaneously quenched to 0 K.[3,14,33] Such simulations should provide the connection between measurements of the structure of quenched water to the structure at the temperatures of interest.

**Methods**

*A. Experimental set-up*

Amorphous solid water (ASW) films were vapor deposited onto the substrate under ultrahigh vacuum (~$10^{-10}$ torr) at 70 K using a calibrated molecular beam source. A Pt(111) single crystal or a graphene layer on Pt(111) were used as substrates.[19,20] The water films were 6 mm in diameter and a few nanometers thick. For the results shown in Figures 1 – 3, the water coverages were 50 monolayers (ML), where 1 ML $\equiv 1.0 \times 10^{19}$ $H_2O/m^2$. The thicknesses of the films were not measured but were presumed to change due to changes in the density of water as a function of temperature. For a nominal density of 1 g/cm$^3$, a 50 ML film is 15 nm thick.

The pulsed heating method has been described in detail previously.[19,20] Briefly, laser pulses from a Nd-YAG laser ($\lambda$ = 1,064 nm, 1 – 5 Hz, pulse width ~10 ns) transiently heat the near-surface region of the substrate and the adsorbed water films to a maximum temperature, $T_{max}$, followed by rapid diffusion of the heat deeper into the substrate resulting in maximum heating and cooling rates of ~2×10$^{10}$ and ~5×10$^9$ K/s, respectively. The desorption of crystalline ice was used to determine $T_{max}$ as function of laser pulse energy. The optical setup was designed to produce a "flat-top" laser pulse profile (i.e. a constant lateral intensity) that had a larger diameter than the water film such that the same temperatures were obtained across the water film. Nonetheless, small deviations from an ideal "flat-top" beam profile led to small lateral variations in $T_{max}$. For example, lateral variations in $T_{max}$ were approximately ± 4 K at $T_{max}$ = 215 K. Calculations show that accounting for the lateral temperature variations for the results shown in Figure 4 results in a slightly narrower range of temperatures (e.g. $\Delta T$ ~ 8.2 K instead of 8.5 K).

Because of the substantial differences in the band shape for isolated HOD in H$_2$O for crystalline ice and liquid water, water with ~10% HOD in H$_2$O was used to increase the sensitivity for detecting the onset of crystallization. Infrared absorption reflection spectroscopy (IRAS) was used to monitor the changes in local structure, as evidenced by spectral changes in the OH stretching region (~3000 – 3700 cm$^{-1}$) and the isolated HOD stretch region (~2350 – 2500 cm$^{-1}$). All the IR spectra were acquired at 70 K after a given number of heat pulses, $N_p$. The typical spectra had 500 scans and were collected at a resolution of 4 cm$^{-1}$. The as-dosed ASW films have IR spectra that are similar to films prepared by hyper-quenching from $T_{max}$ = 297 K (i.e. "HTL"). For the results presented here, the as-dosed films were always converted to either LTL or HTL to begin the experiments. However, for other experiments in which the as-dosed or HTL films were used as the initial configuration, the films obtained the same steady-state configurations with similar kinetics.

### B. Procedure for determining $f_{LTL}$, $f_{HTL}$, and $f_{CI}$ from the IR spectra

For water heated to $T = T_{max}$ for $N_p$ heat pulses, $f_{LTL}(T, N_p)$, $f_{HTL}(T, N_p)$, and $f_{CI}(T, N_p)$ were determined by fitting the IR spectrum to a linear combination of reference spectra for LTL, HTL and CI. As described in the main text, the "low-temperature liquid," LTL, was prepared by annealing the water at 135 K for 130 s. Note that changing the annealing temperature by $\pm 5$ K, or the annealing time by $\pm 100$ s, led to essentially indistinguishable spectra for the LTL. The "high-temperature liquid," HTL, was prepared by transiently-heating a film to $T_{max} = 297$ K, typically for 3 heat pulses. As seen in Figure 4, any pulsed heating temperature above ~250 K produces water films with the same HTL structure as indicated by the indistinguishable IR spectra. The CI spectra were obtained from films that had been crystallized by pulsed heating. For the results shown in the main text, the crystalline fraction, $f_{CI}$, was small (i.e. $\leq 0.01$ except for $N_p > 500$ in Fig. 3b). In that case, $f_{LTL}$ and $f_{HTL}$ are simply related to each other: $f_{LTL} + f_{HTL} \approx 1$. This is consistent with the observation that the spectra display isosbestic points prior to the onset of crystallization. Extended Data Fig. 4 shows several spectra along with the results of the fitting procedure. In these examples, the crystalline component of the fitting was $\leq 0.01$ and is omitted for clarity. The top and middle rows of the figure show several "steady-state" spectra (green lines), the results of the fitting procedure (black dashed lined), and the scaled LTL (blue line) and HTL (red line) components. The bottom row in Extended Data Fig. 4 shows several spectra and the resulting fit for a water film that was initially HTL that was heated to 190 K after various numbers of heat pulses.

The "steady-state" value of $f_{HTL}(T)$ at a given temperature was determined by two methods. In the first approach, $f_{HTL}(T)_i$ for an experiment (where the subscript $i$ refers to an individual experiment) was determined from the average of the values of $f_{HTL}(T, N_p)$ for several spectra just prior to crystallization. $f_{HTL}(T)$ was then found by averaging over all the individual experiments at a given temperature: $f_{HTL}(T) = \frac{1}{N}\sum_{i=1}^{N} f_{HTL}(T)_i$. In the second approach, $f_{HTL}(T, N_p)$ prior to crystallization was fit to a stretched exponential function and $f_{HTL}(T)_i$ was taken as $f_{HTL}(T)_i = f_{HTL}(T, N_p \rightarrow \infty)$. The values of $f_{HTL}(T)$ determined from fitting $f_{HTL}(T, N_p)$ to a stretched exponential function are the same as the values obtained directly from the IR spectra just prior to the onset of crystallization within the experimental uncertainty (see error bars in Figure 4).

**Acknowledgements**

This work was supported by the U.S. Department of Energy (DOE), Office of Science, Office of Basic Energy Sciences, Division of Chemical Sciences, Geosciences, and Biosciences. This research was performed using EMSL, a national scientific user facility sponsored by DOE's Office of Biological and Environmental Research and located at the Pacific Northwest National Laboratory, which is operated by Battelle for the DOE.

**Author contributions**

The research was designed and supervised by GAK and BDK. LK and WAT conducted the experiments and analyzed the data. LK and GAK wrote the manuscript, with input from all authors.

**Data Availability**

The datasets generated and/or analyzed during the current study are available from the corresponding author on reasonable request.

**Competing Interests**

The authors declare no competing interests.

**Correspondence and requests for materials** should be addressed to GAK.

**Figures**

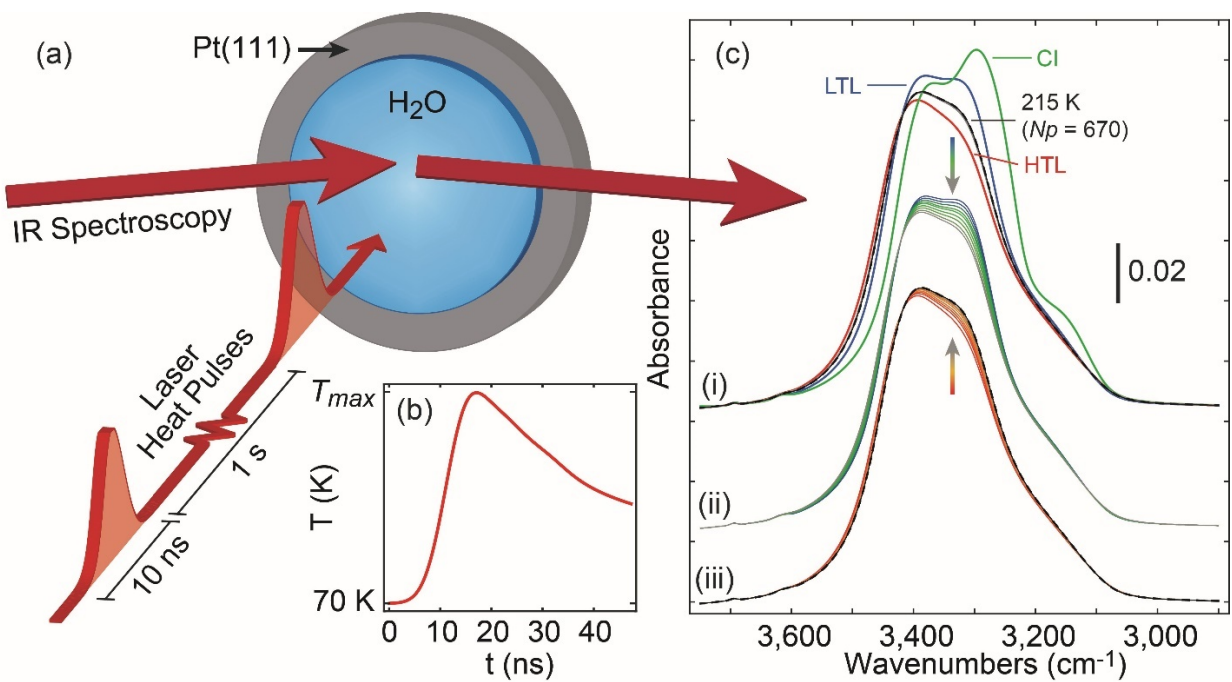

**Figure 1.** (a) Schematic showing the experimental approach for transiently-heating nanoscale water films adsorbed on a Pt(111) (or graphene/Pt(111)) substrate using nanosecond laser pulses. (b) Calculated water temperature versus time, $T(t)$, during a heat pulse. The maximum temperature, $T_{max}$, is determined by the laser pulse energy. (c) Infrared reflection absorption spectra of transiently-heated, 50 monolayer water films on graphene/Pt(111). The IR spectra were collected at 70 K. (c)(i) Spectra for water quenched from $T_{max} = 297$ K (HTL, red line), annealed at 135 K (LTL, blue line), and crystalline ice (CI, green line) are all distinct. Starting with either HTL or LTL and heating to 215 K for 670 heat pulses, the resulting spectra are essentially identical (black and gray lines). IR spectra for water heated to 215 K for (c)(ii) LTL(initial) after $N_p = 4$, 10, 19, 31, 55, 100, 210 and 670, and (c)(iii) HTL(initial) after $N_p = 1$, 2, 4, 7, 31, 100, and 670.

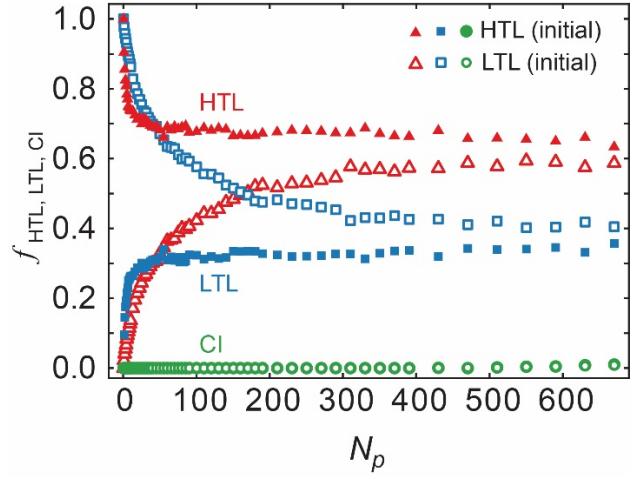

**Figure 2**. $f_{HTL}$, $f_{LTL}$, and $f_{CI}$ versus $N_p$ for water heated to 215 K. Starting with either HTL or LTL, the films relax to a steady-state configuration with $f_{HTL} = 0.62 \pm 0.03$ prior to crystallization. For $N_p < 700$, $f_{CI} \leq 0.01$ (green circles).

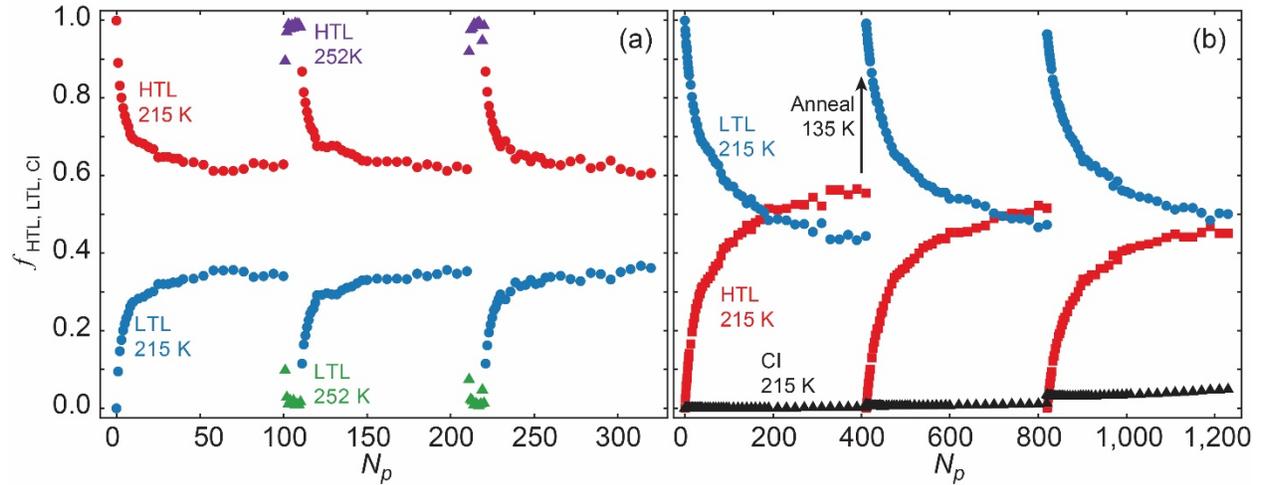

**Figure 3.** $f_{HTL}$ and $f_{LTL}$ versus $N_p$ for water alternately heated between (a) 215 K and 252 K or (b) 215 K and 135 K. (a) Each time water was transiently-heated at $T_{max}$ = 215 K, $f_{HTL}$ (red circles) decreased from 1 to ~0.62, while pulsed heating at $T_{max}$ = 252 K, restored $f_{HTL}$ (purple triangles) to ~1 within a couple of pulses. For this experiment, $f_{CI}$ was less than 0.01, such that $f_{LTL}$ ~ 1 – $f_{HTL}$ (blue circles – 215 K, and green triangles – 215 K). (b) Starting from LTL, $f_{LTL}$ decreased to ~0.4 after 400 heat pulses at 215 K (blue circles). Upon annealing at 135 K, $f_{LTL}$ increased to 1. For this experiment, the onset of crystallization ($f_{CI}$ ~ 0.01, black triangles) was observed at $N_p$ ~ 500. However, the liquid water component of the film continued the reversible structural transformations even for $N_p$ > 500. In both (a) and (b), the structural transformations of deeply supercooled water are not due to (irreversible) changes associated with crystallization or with crossing a spinodal. The corresponding IR spectra are shown in Extended Data Fig. 5.

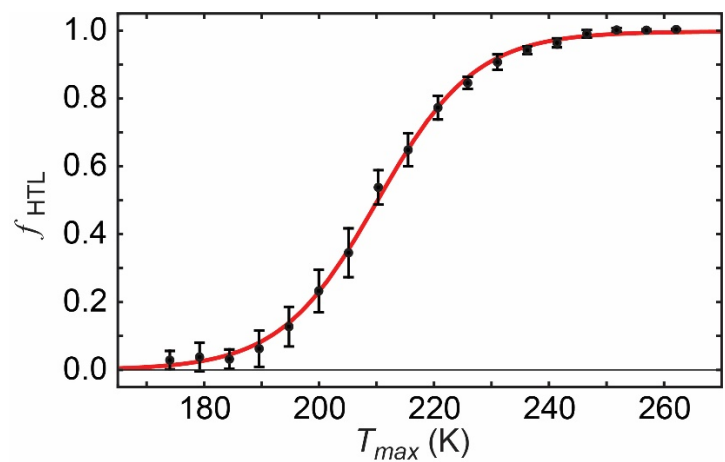

**Figure 4**. The steady-state fraction of water that corresponds to HTL, $f_{HTL}$, versus the pulsed-heating temperature (black circles). The red line is a fit to the data using a logistic function. $f_{HTL}$ is ~1 for $T_{max} > 245$ K due to kinetic limitations that make the experiments insensitive to the structure of water at higher temperatures (see text). The error bars show 1 standard deviation for the measured values.

# Reversible structural transformations in supercooled water from 135 to 245 K

Loni Kringle†, Wyatt A. Thornley†, Bruce D. Kay*, and Greg A. Kimmel*

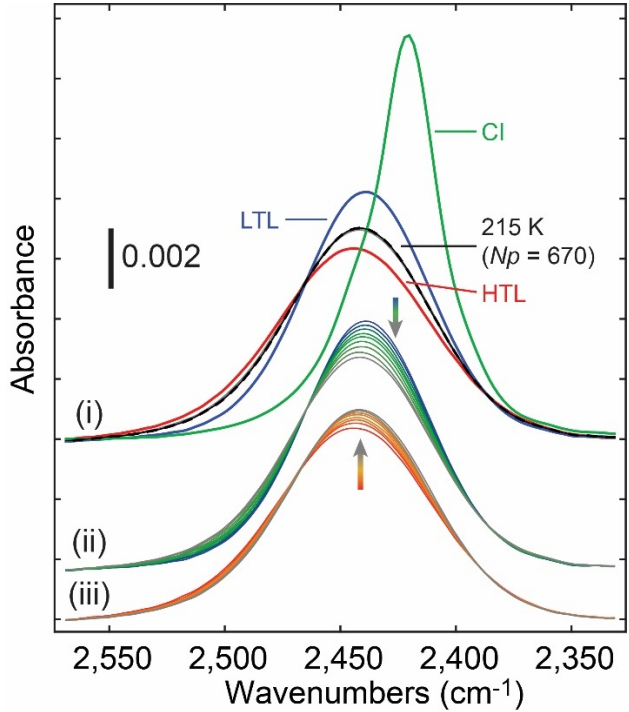

**Extended Data Figure 1.** IR spectra in the isolated HOD stretch region. The spectra shown are the same as those in Figure 1b, but here focusing on the behavior of the isolated HOD molecules within the water (10% of the molecules are HOD). (i) The IR spectra for HTL, LTL and CI (red, blue and green lines, respectively) are all distinct. After 670 heat pulses, the resulting spectra (solid gray and dashed black lines) are independent of the initial configuration and intermediate between the HTL and LTL spectra. (ii) and (iii): Water initially prepared as either LTL (ii) or HTL (iii) that was heated to $T_{max} = 215$ K evolved to the same "steady-state" structure that was intermediate between the HTL and LTL spectra. For $N_p <$ 700, the fraction of crystalline ice within the films was less than 1%. Fitting the IR spectra in the isolated HOD stretch region to a linear combination of HTL, LTL and CI spectra yields results that are the same within experimental uncertainty to those shown in Figure 2.

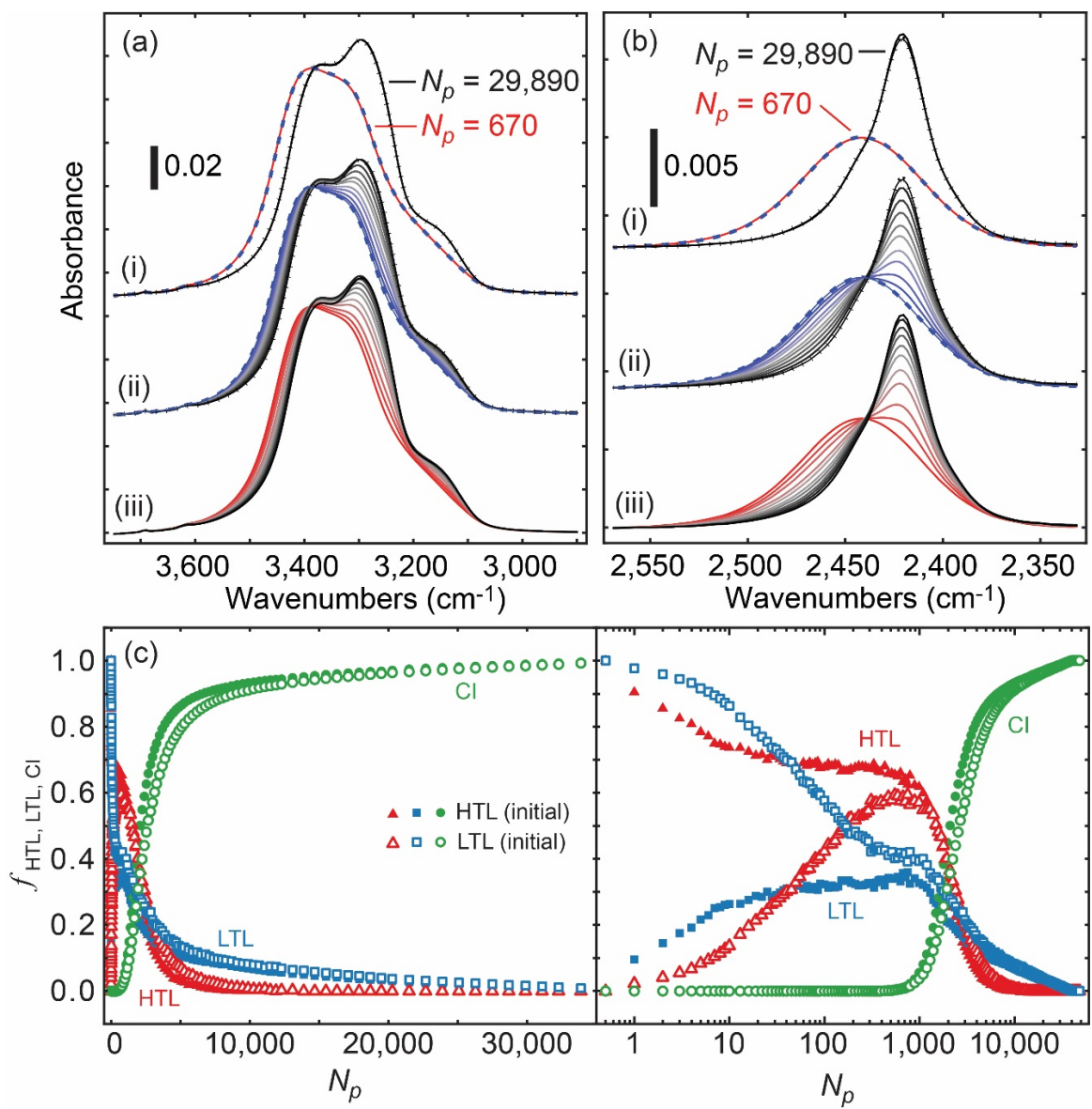

**Extended Data Figure 2.** Crystallization of water heated to $T_{max}$ = 215 K. For the experiments shown in Figure 1, the water eventually crystallizes (i.e. $f_{CI}$ > 0.01 for $N_p$ > 670). Panels (a) and (b) show the corresponding IR spectra in the OH-stretch and isolated OD-stretch regions, respectively. (a, b)(i): IR spectra for $N_p$ = 670 (dashed blue and solid red lines) and 29890 (black line or black line with ×'s) are the same for water films with initial configurations of LTL or HTL. (a)(ii, iii) and (b)(ii, iii): Spectra are shown for $N_p$ = 670, 1290, 1590, 1940, 2240, 2690, 3140, 3890, 5090, 8690 and 29890. As the fraction of crystalline ice increases, the spectra red shift in both OH-stretch and isolated OD-stretch regions. (a, b)(ii): Initial configuration was LTL. (a, b)(iii): Initial configuration was LTL. (c) $f_{HTL}$, $f_{LTL}$ and $f_{CI}$ versus $N_p$ determined from fitting the spectra to a linear combination of LTL, HTL and CI. The left and right panels show the results versus $N_p$ on a linear and logarithmic scale, respectively.

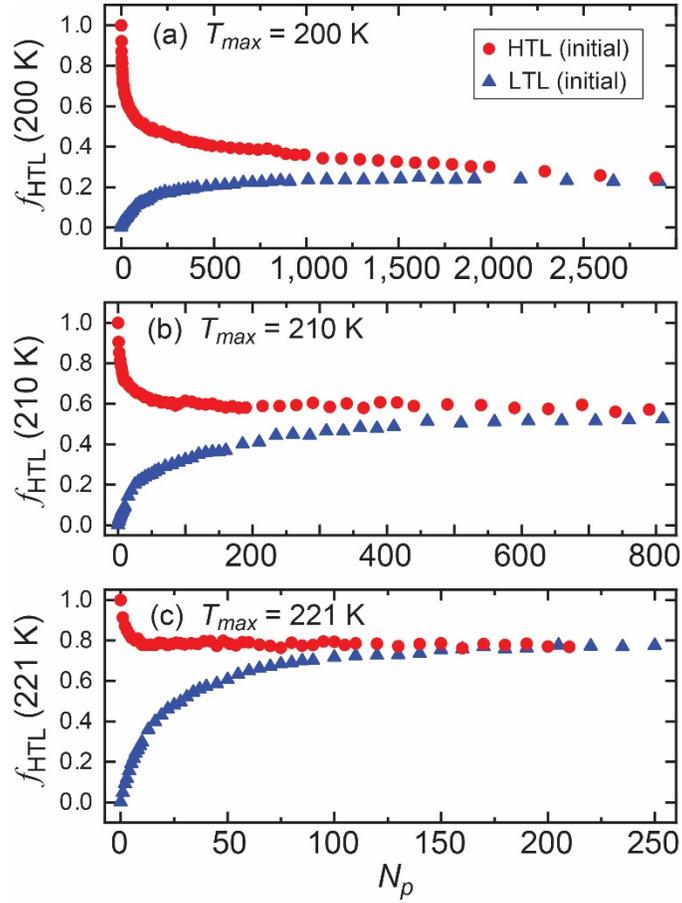

**Extended Data Figure 3**. $f_{HTL}$ versus the number of heat pulses, $N_p$, for water that was transiently-heated to (a) $T_{max} = 200$ K, (b) $T_{max} = 210$ K, (c) $T_{max} = 221$ K. For each temperature, the relaxation kinetics were measured for two different starting structures for the water: HTL (red circles) or LTL (blue triangles). For a given $T_{max}$, the two water films approach the same steady-state structure, although the kinetics are consistently slower when starting from the LTL state. In each case, $f_{Cl} \leq 0.01$ for the range shown.

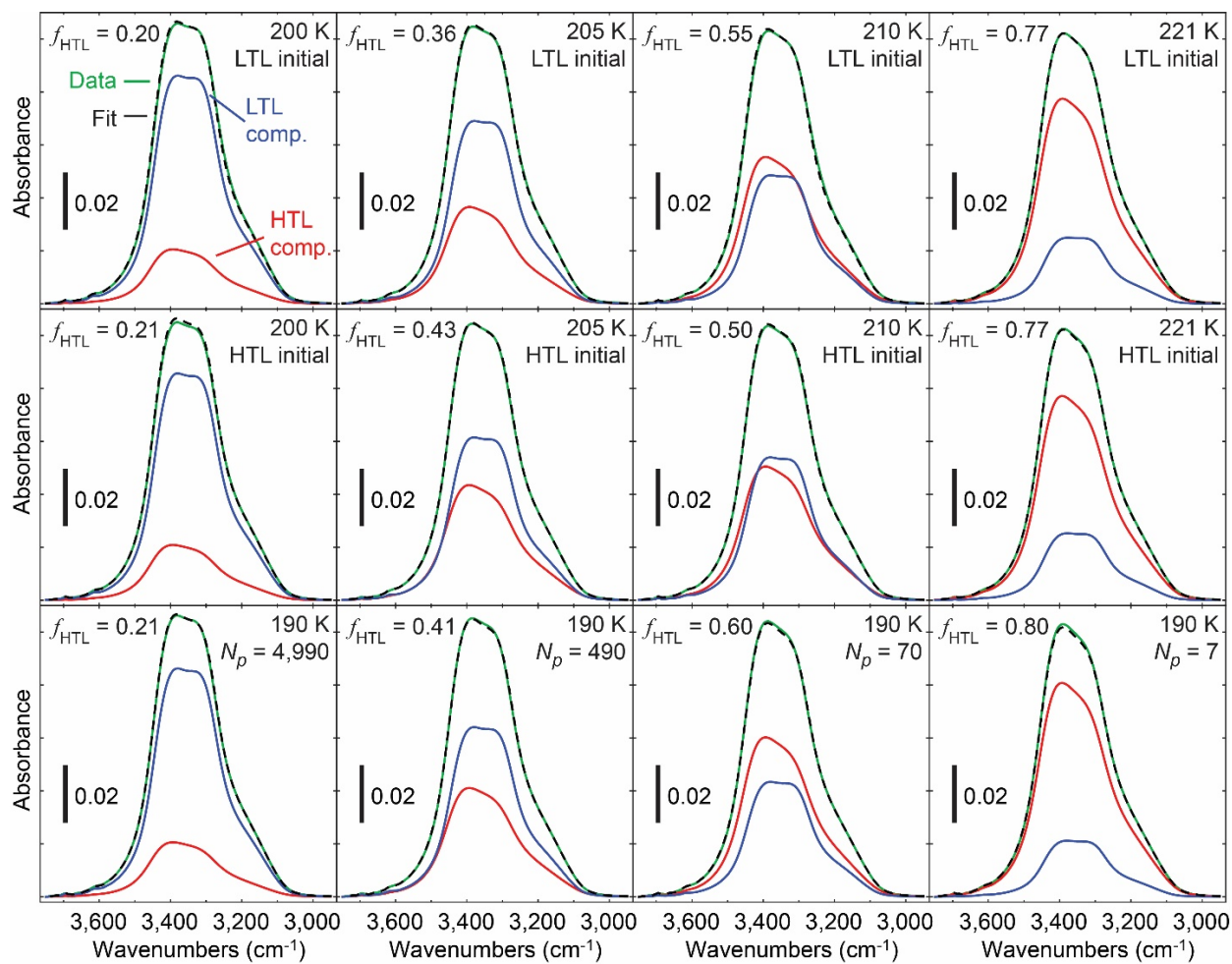

**Extended Data Figure 4**. IR spectra of ~15 nm thick (50 ML) water films transiently-heated to different temperatures and the results of fitting them to linear combinations of LTL, HTL, and crystalline ice. The green lines show the data, the dashed black lines are the fits to the data, and the blue and red lines are the corresponding LTL and HTL components. For these results, the fraction of crystalline ice, $f_{CI}$, was negligible ($\leq 0.01$) so it was not displayed in the figure. The top row (initial structure = HTL) and middle row (initial structure = LTL) show the steady-state spectra for water at the indicated temperatures and compositions. The structure just prior to the onset of measurable crystallization ($f_{CI} \sim 0.01$) was defined as the steady-state structure. The bottom row shows the spectra for a water film that was initially converted to HTL and then heated to 190 K for $N_p$ = 7, 70, 490 and 4990 pulses (right to left in the figure). Because the steady-state value of $f_{HTL}$ is ~0.06 at 190 K (see Fig. 4), these spectra reflect the water structure prior to reaching steady-state. These spectra that were obtained as the film evolved towards the steady-state structure can also be fit as a linear combination of HTL and LTL.

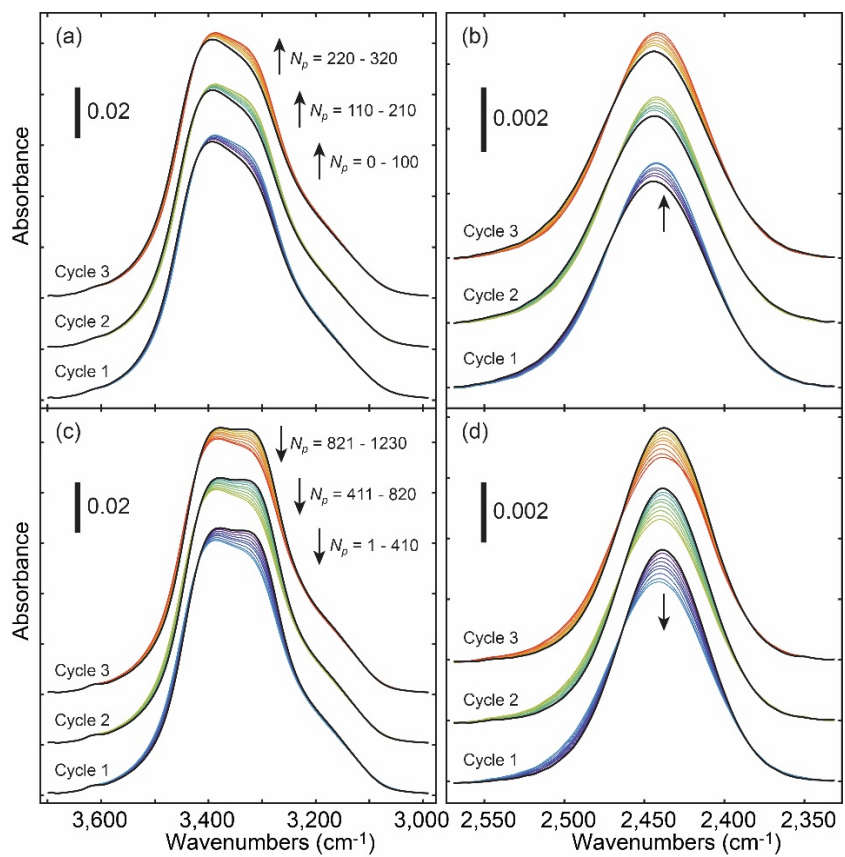

**Extended Data Figure 5**. IR spectra for the results shown in Figure 3 in the OH-stretch region ((a) and (c)) and the isolated HOD region ((b) and (d)). (a) and (b): Starting from HTL (Cycle 1, black line), the water was heated to $T_{max}$ = 215 K (for 100 pulses) and then $T_{max}$ = 252 K (for 10 pulses). This sequence was repeated two more times (Cycles 2 and 3). The spectral changes upon heating to 215 are shown for each cycle, while the rapid conversion of the structure back to HTL upon heating to 252 K is not shown. As the number of heat pulses at 215 K increases in each cycle, the spectra change due to the increasing fraction of LTL. (c) and (d): Starting from LTL (Cycle 1, black line), the water was heated to $T_{max}$ = 215 K (for 400 pulses). As the number of heat pulses increases, the spectra change due to the increasing fraction of HTL (Cycle 1, blue – purple lines). After 400 heat pulses, the film was heated isothermally at 135 K for 130 s and LTL was recovered (Cycle 2, black line). Pulsed heating to $T_{max}$ = 215 K a second time (Cycle 2, blue – green lines) resulted in similar structural changes as those observed in Cycle 1. Another round of annealing at 135 K and pulsed heating to 215 K (Cycle 3, red lines) also produced similar changes. Although it is difficult to see in the spectra, the crystalline fraction has increased to ~0.01 at the beginning of the second cycle of transient heating to 215 K.

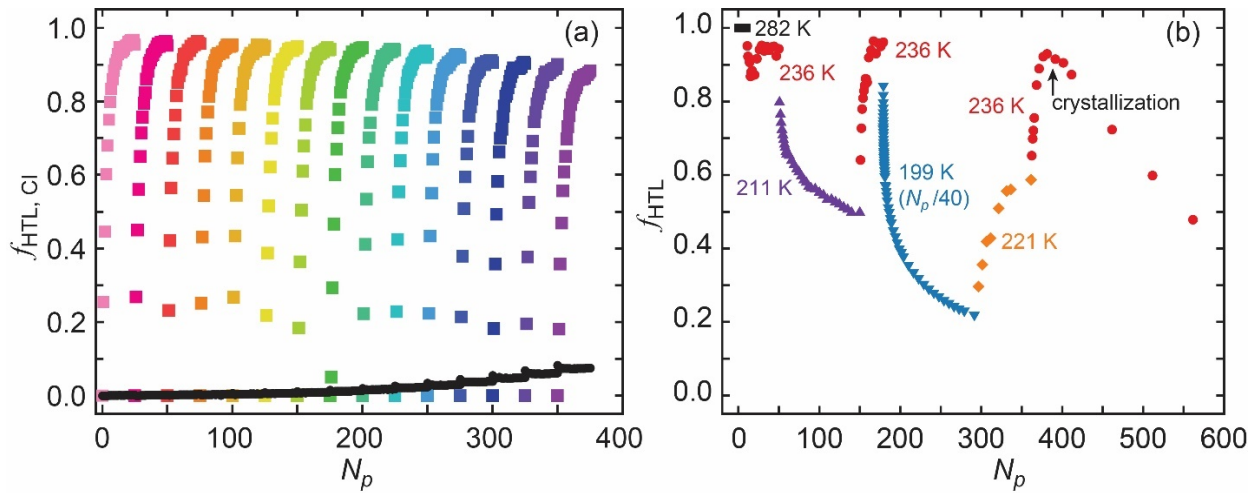

**Extended Data Figure 6**. Reversible structural changes for supercooled water. (a) $f_{HTL}$ and $f_{CI}$ versus $N_p$ for a water film where the temperature was cycled 15 times between $T_{max} = 241$ K (25 heat pulses for each cycle) and isothermal annealing at 135 K for 130 s. The different colored symbols show $f_{HTL}$ versus the total number of heat pulses. After 25 heat pulses for each cycle at 241 K, the film was isothermally annealed at 135 K and $f_{HTL}$ resets to zero. The black circles show $f_{CI}$ as the (non-reversible) crystalline phase slowly grows as the experiment proceeds. Note that the structure of the liquid component continues to reversibly respond to changes in the temperature even as the crystalline component increases. (b) $f_{HTL}$ versus $N_p$ for a water film heated to several different maximum temperatures. An HTL film was prepared by hyper-quenching at $T_{max} = 297$ K. It was then heated to $T_{max} = 282$ K (black squares), 236 K (red circles), and 211 K (purple triangles) for which $f_{HTL}$ dropped to ~0.5, but the film had not reached "steady state." Upon heating to 236 K a second time (red circles), $f_{HTL}$ returns to ~0.95 which is the value it had on the first cycle at 236 K. Heating to a lower temperature,199 K (blue diamonds, $Np$/40), led to another decrease in $f_{HTL}$, before the temperature was increased to 221 K (orange diamonds). Finally, during the third cycle of heating the water to 236 K (red circles), the film began to crystallize causing $f_{HTL}$ to decrease, this time without a corresponding increase in $f_{LTL}$ (not shown).

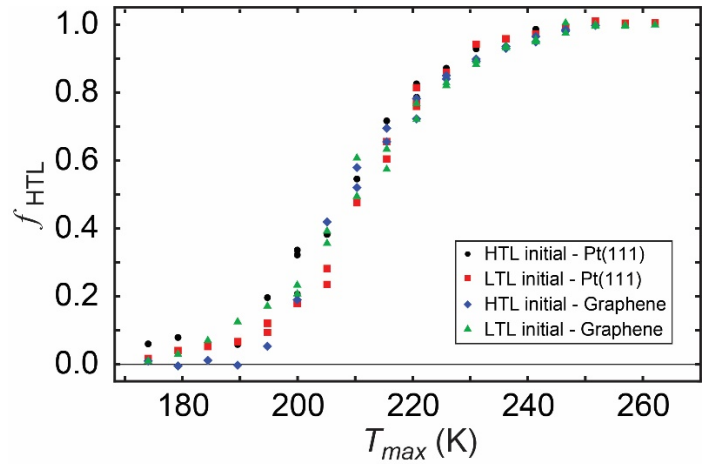

**Extended Data Figure 7**. The steady-state fraction of HTL, $f_{HTL}$, versus temperature. The experiments were conducted for four different configurations: The water films were deposited on Pt(111) (black circles and red squares) or graphene/Pt(111) (blue diamonds and green triangles). Prior to transiently-heating the water to $T_{max}$, the initial structure for the water was either HTL (black circles and blue diamonds) or LTL (red squares and green triangles). The average of these individual experiments is reported in Figure 4 with the error bars showing the standard deviation of the individual measurements.

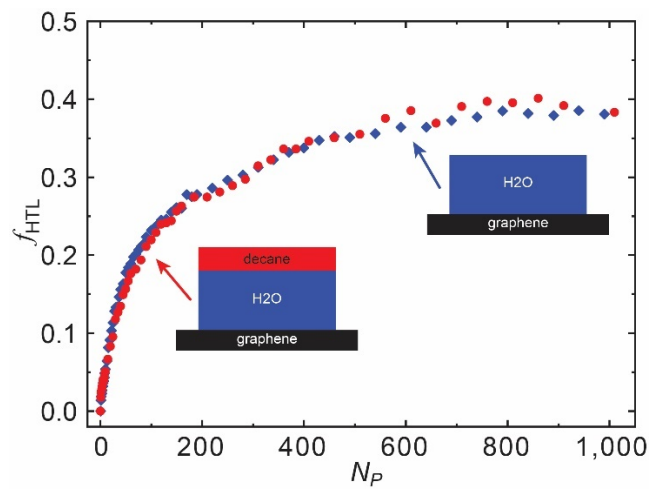

**Extended Data Figure 8.** $f_{HTL}$ versus $N_p$ for $T_{max} = 205$ K for a water film with and without a 20 ML decane layer adsorbed on top. Changing the outer interface for the water film did not have an appreciable effect on the relaxation kinetics or the steady-state structure.

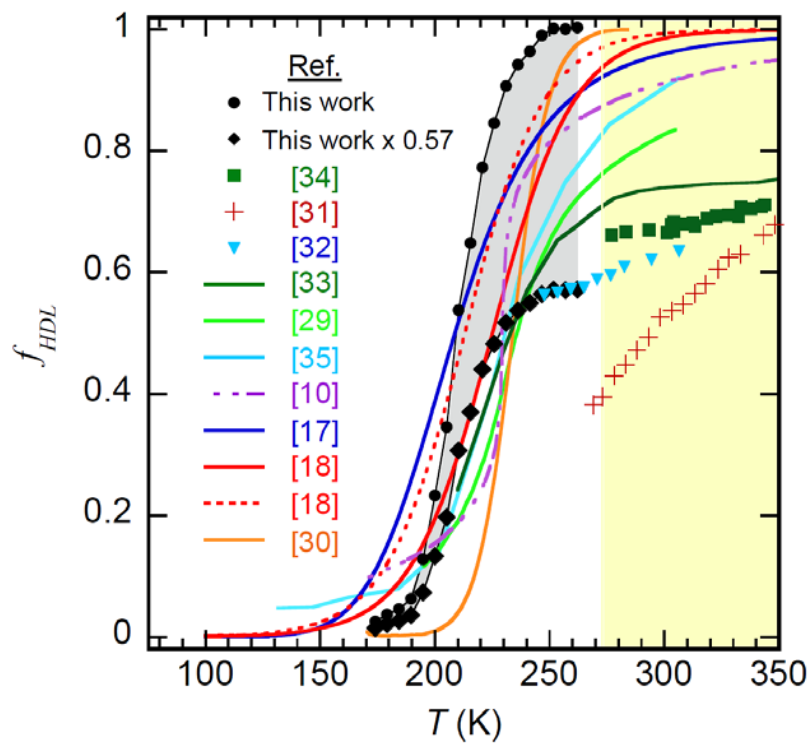

**Extended Data Figure 9.** $f_{HDL}$ versus temperature. The filled black circles show the current measurements with assumption that the HTL spectra corresponds to $f_{HDL} = 1$. The black diamonds show the same data assuming the HTL spectra have $f_{HDL} \sim 0.57$. The gray shaded region shows the likely range for $f_{HDL}$ that is consistent with our data. The lower boundary of the shaded region was chosen such that the current results match Raman spectroscopy measurements of supercooled water (blue triangles).[32] The yellow shaded region corresponds to T > 273 K. The green squares and red crosses show X-ray spectroscopy[34] and IR[31] measurements for ambient water, respectively. There have been several attempts to fit experimental data to models assuming that water is a mixture of two liquids.[3,10,29] The dashed purple[10] and light green lines[29] show two examples. Hestand and Skinner analyzed experimental data by assuming that the observables had contributions of the "pure" HDL and LDL components weighted by their mole fraction in the water at any temperature.[18] The solid red and dashed red lines show the results of this analysis for X-ray scattering and self-diffusion data, respectively. A different two-state analysis of the diffusion data led to similar results (dark blue line).[17] The orange line shows another two-state analysis using density data versus temperature and pressure.[30] Results of classical molecular dynamics simulations of normal and supercooled water have also been analyzed in terms of the HDL-LDL framework. The light blue[35] and dark green lines[33] show two examples using TIP4P/2005.